\documentclass[preprint2]{aastex}
\usepackage{psfig}
\pssilent
\begin{document}
\title{ Multiplicity of X-Ray Selected T Tauri Stars in Chamaeleon%
\footnote{Based on observations obtained at the European South\-ern
Observatory, La Silla, proposal number 56.E-0197}}
\author{ Rainer K\"ohler}
\affil{%
	Center for Astrophysics and Space Sciences,
	University of California San Diego,
	Mail Code 0424, Gilman Drive 9500, La Jolla, CA 92093-0424, USA
}
\email{rkoehler@ucsd.edu}
\begin{abstract}
We report on a multiplicity survey of a sample of X-ray selected young
stars in the Chamaeleon association.  We used speckle-interferometry
and direct imaging to find companions in the separation range $0.13''$
to $6''$.  After correction for chance alignment with background
stars, we find a multiplicity (number of binaries or multiples divided
by number of systems) of $(14.0\pm 4.3)\,\%$ and a companion star
frequency (number of companions divided by number of systems) of
$(14.7\pm 5.1)\,\%$.  Compared to solar-type main-sequence stars, the
companion star frequency is lower by a factor of $0.61\pm 0.27$.  This is
remarkably different from the high multiplicity found in the
Taurus-Auriga star-forming region and for T~Tauri stars in Chamaeleon
known before ROSAT.  We find only a few binaries with projected
separations of more than 70\,AU, also in contrast to the results for
stars known before ROSAT.  This indicates that the X-ray selected
stars belong to a different population than the stars known before
ROSAT, a hypothesis further supported by their Hipparcos distances and
proper motions.
\end{abstract}
\keywords{stars: pre-main-sequence -- binaries: visual -- infrared:
stars -- surveys -- techniques: interferometric}
\section{Introduction}
Recent studies of multiplicity in several star-forming regions have
shown that multiplicity is -- besides the initial mass function -- one
of the key characteristics of the star-formation process.  Surveys of
T Tauri stars (TTS) in the T association Taurus-Auriga
\citep{Leinert93,Ghez93,Koehler98} found an overabundance of binaries
compared to solar-type main-sequence stars \citep{D+M91} by a factor
of two.  On the other hand, surveys of low-mass stars in the Orion
Trapezium Cluster \citep{Prosser94,Padgett97,Petr98} found a
multiplicity that is comparable to or even lower than among
main-sequence stars.  \citet{Scally99} used proper motion data of
stars in the Trapezium cluster to search for binaries with separations
in the range 1000 to 5000~AU.  Their statistical analysis shows that
the cluster contains no binaries at all in this separation range.
Studies of other clusters (\citealp[IC348:][]{Duchene99};
\citealp[Pleiades:][]{Bouvier97}) also found a binary frequency
comparable to main-sequence stars.
An explanation for these results is the disruption of binaries in
stellar encounters, which happen predominantly in dense clusters
\citep{Kroupa95b}.  However, it is also possible that the multiplicity
depends on environmental parameters like magnetic fields or the
cloud temperature \citep[e.g.][]{Durisen94}.

\begin{figure*}[ht]
\centerline{\psfig{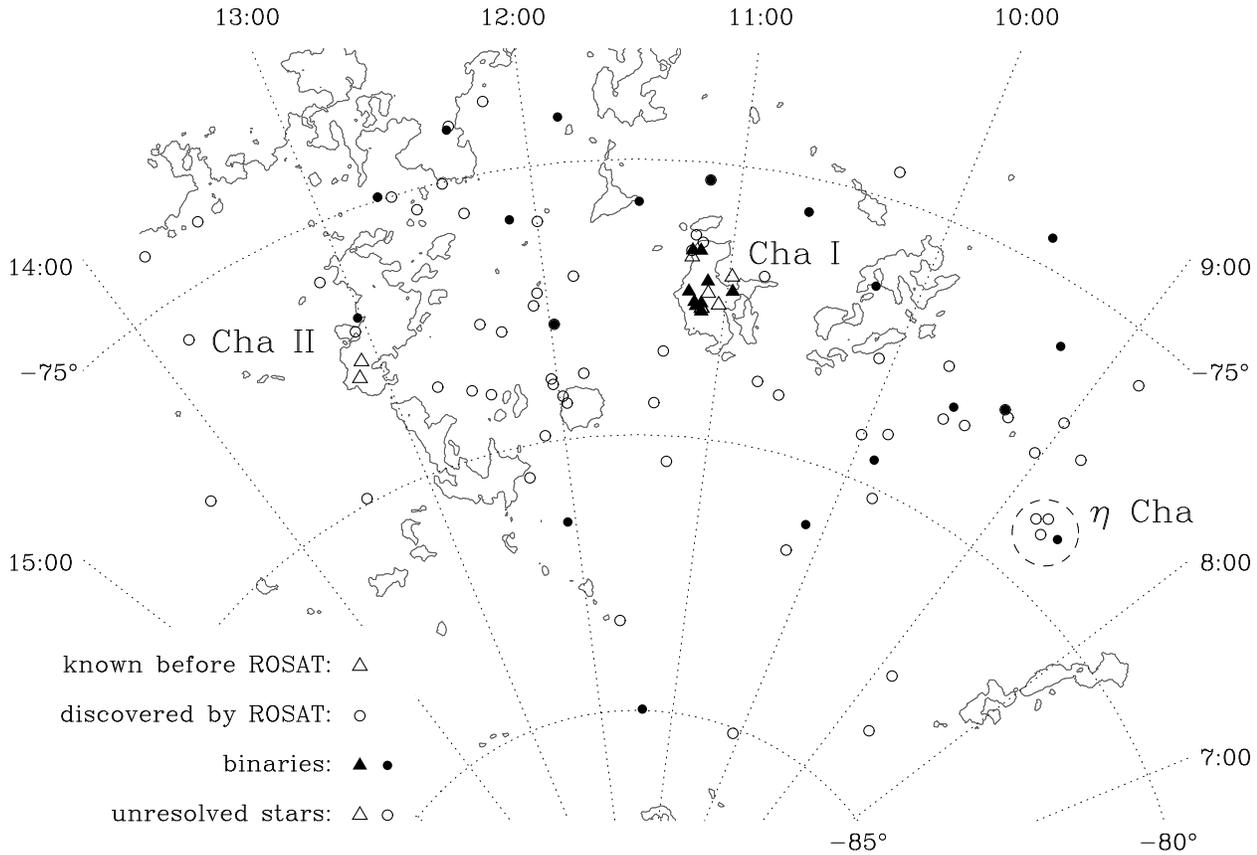}}
\caption{Spatial distribution of the stars in our sample and the
sample of \citet{Ghez97}. The contours mark the emission found in the
IRAS $100\rm\,\mu m$ survey.  Cha I and II are the cloud complexes
where most TTS known before ROSAT are located.  The large dashed
circle encloses the four stars in the $\eta$ Chamaeleontis cluster
\citep{Mamajek99}.}
\label{MapFig}
\end{figure*}

\citet{Kroupa95a} proposed to use the binary frequency and their
distribution of periods to determine the dominant mode of star
formation, i.e. if most stars form in dense regions like the Orion
Cluster, or in loose associations like Taurus-Auriga (``inverse
population synthesis'').  However, although Taurus-Auriga is the best
studied T~association, it is by no means clear that it is a typical
example with respect to the formation of binaries.  Some authors
surveyed other star-forming regions
\citep{ReipurthZinn93,Brandner96,Ghez97}, but the number of stars
observed or the spatial resolution was very limited, which causes
large statistical errors.  Furthermore, it turned out that even within
one star-forming region (Scorpius-Centaurus), the period distribution
of binaries in two subregions can be significantly different
\citep{B+K98,Koehler2000}.  We can expect to find even larger differences
between different star-forming regions.

In the light of these results, it is desirable to conduct a survey of
a large number of stars in other T associations, in order to find
out if Taurus-Auriga is a typical example or the exception to the
rule.  Thanks to the ROSAT All-Sky Survey, sufficiently large samples
of TTS are available for the T associations Chamaeleon and Lupus.  We
carried out multiplicity surveys for both of them.  This paper
describes the survey in Chamaeleon, Lupus will be presented elsewhere.

\citet{Ghez97} already searched for binaries among TTS in Chamaeleon.
However, only the observations of 17 stars were sensitive enough to
give meaningful results.  Furthermore, they observed only stars known
before ROSAT, i.e. stars located on or near the dark clouds and mainly
Classical T Tauri stars (CTTS).  The results of the ROSAT All-Sky
survey have shown that pre-main-sequence stars can be found in much
larger areas around star-forming regions, and that the majority of
pre-main-sequence stars are Weak-line T Tauri stars
(WTTS; \citealt{Walter88,Neuhaeuser95}).

Our survey encompasses a list of WTTS found with the help of ROSAT.
Section 2 describes the object list and section 3 our observations
and data reduction.  The results are presented and discussed in
section 3 and 4, while section 5 contains the summary and
conclusions.

\section{The Sample}
Our survey is based on the results of \citet{Alcala95} and
\citet{Covino97}.  Alcal\'a et al.\ did follow-up observations of
optical counterparts of X-ray sources in the Chamaeleon region
detected in the ROSAT All-Sky Survey. They identified 82 young
low-mass stars based on the presence of the Li~$\lambda6707$
absorption line and the spectral type.  A complete list with
coordinates, photometric data, and finding charts can be found in
\citet{Alcala95}.  Fig.\ \ref{MapFig} shows the spatial distribution
of these stars and the stars observed by \citet{Ghez97}.

The four stars around $\rm8^h42^m$, $-79^\circ$ are probably members
of the $\eta$ Chamaeleontis cluster \citep{Mamajek99}.  Our results
for these stars are listed in Table~\ref{BinTab} and \ref{SingTab},
but we do not use them for the discussion of the binary properties in
the Chamaeleon star-forming region.  The binaries in $\eta$ Cha will
be discussed in \citet{Koehler01}.  Two sources
(\objectname{RXJ~1039.5-7538}N+S) are separated by only $5''$,
therefore we count them as one binary.  This leaves us with a sample
of $82-4-1=77$ systems.

It has been questioned recently \citep{Briceno97} whether stars found
with low-resolution spectroscopy are indeed pre-main-sequence (PMS)
objects or if they are somewhat older stars which have already reached
the zero-age main-sequence (ZAMS).  In order to unambiguously
distinguish PMS and ZAMS objects, \citet{Covino97} carried out
observations with high spectral resolution of some 70 stars of the
sample. They compared the equivalent width of the Li~$\lambda6707$
line to that of Pleiades stars of the same spectral type and
classified the stars as PMS or ZAMS.  They found that more than 50\,\%
of the stars could be confirmed to be PMS objects.

\section{Observations and Data Reduction}

\begin{figure}[ht]
\centerline{\psfig{figure=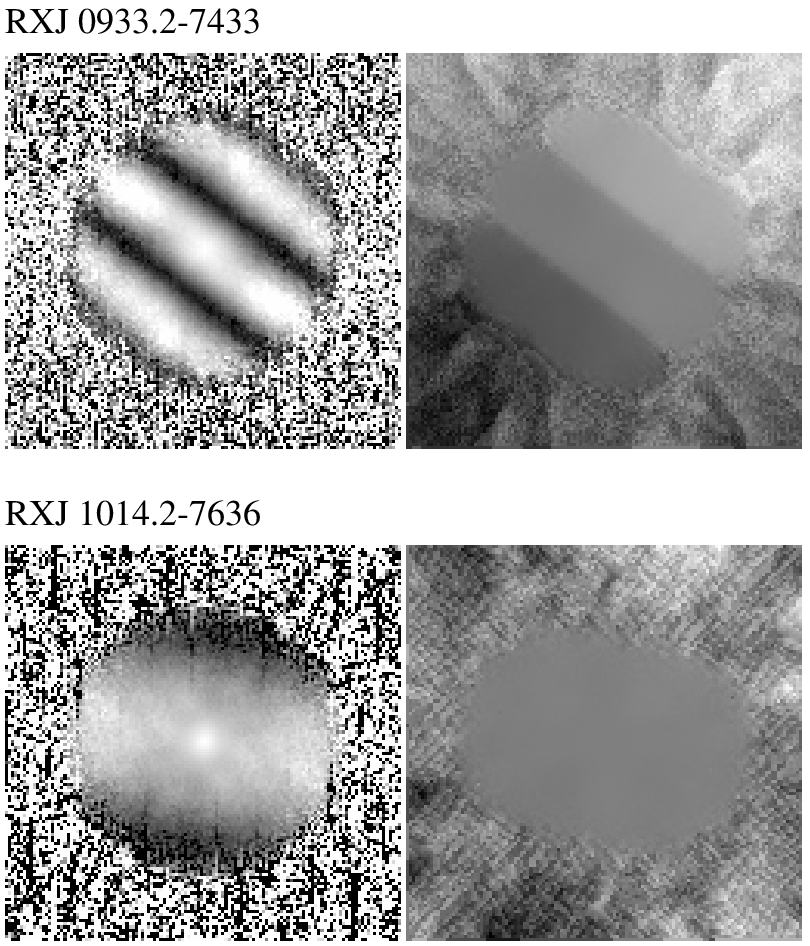,width=\hsize}}%
\caption{Examples of our speckle data.  The images in the first row
show modulus (left) and phase (right) of the complex visibility of
RXJ~0933.2-7433, a close, but well-resolved binary.  The central
maximum and the first maximum on both sides are clearly detected.  The
second row shows modulus and phase of RXJ~1014.2-7636, a potential
binary closer than the diffraction limit of the telescope.  In this
case, we see only the central maximum and the first minimum on each
side.}
\label{ExmplFig}
\end{figure}

The speckle observations were carried out at the ESO 3.5\,m New
Technology Telescope (NTT) on La Silla, Chile, from 29.\,March to
2.\,April 1996.  We used the SHARP~I camera (System for High Angular
Resolution Pictures) of the Max-Planck-Institut for Extraterrestrial
Physics \citep{SHARP}.  All observations were done in the K-band at
$2.2\rm\,\mu m$.

Although speckle interferometry can be considered by now a standard
technique \citep{Leinert92}, no program for speckle data reduction was
publicly available at the time this survey was started.  Therefore we
used our {\tt speckle} program\footnote{Now available on the software
web page of the Center for Adaptive Optics at\\
\url{http://www.ucolick.org/\~{}cfao/distributedsw/ds.html}
and\hfill\\ \url{http://babcock.ucsd.edu/cfao\_ucsd/software.html}},
which was already used for the surveys in Taurus-Auriga
\citep{Koehler98} and Scorpius-Centaurus \citep{Koehler2000}.  In this
program, the modulus of the complex visibility (i.e.\ the Fourier
transform of the object brightness distribution) is determined from
power spectrum analysis, the phase is computed using the Knox-Thompson
algorithm \citep{KnoxThomp74}, and from the bispectrum
\citep{Lohmann83}.  Figure~\ref{ExmplFig} shows examples of
reconstructed complex visibilities.  For a more detailed description
see \citet{Koehler2000}.

\begin{figure*}[ht]
\centerline{\psfig{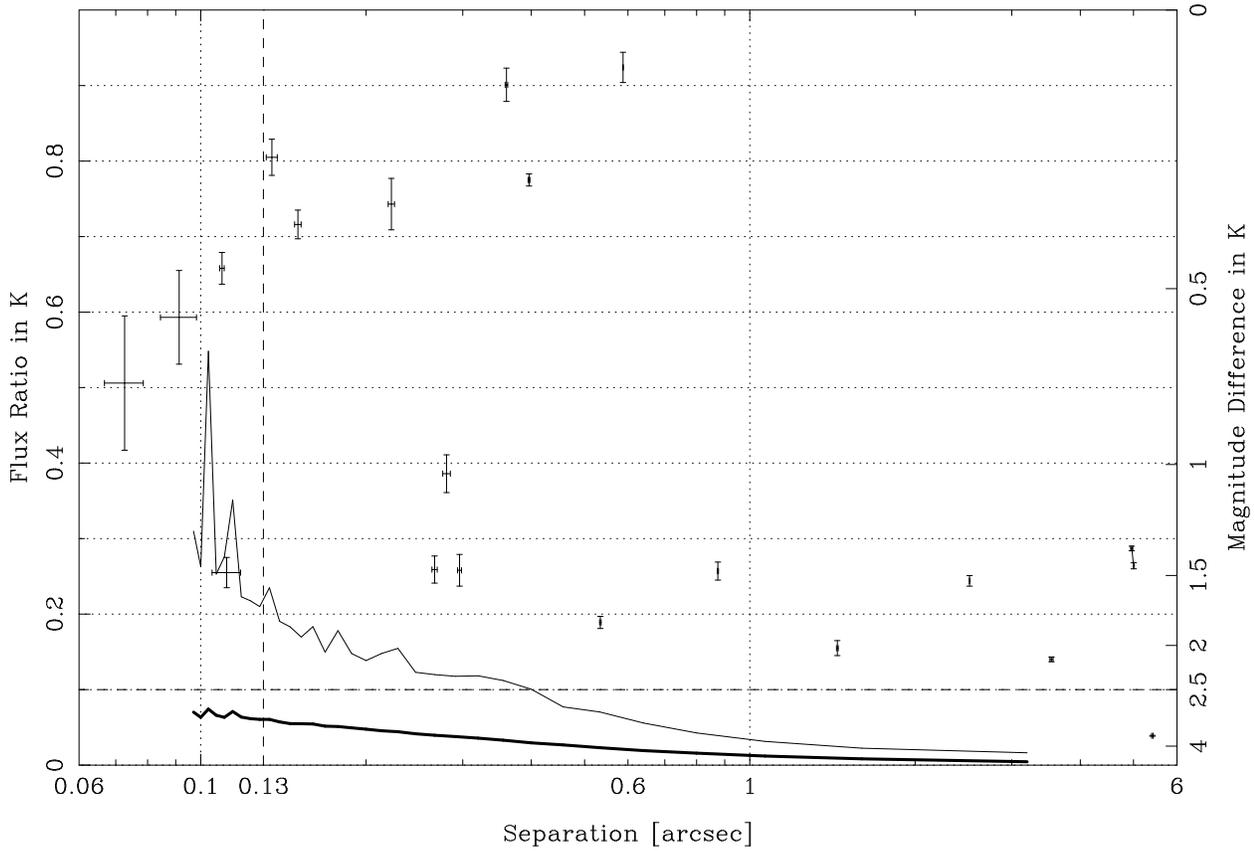}}
\caption{The results of our multiplicity survey in a plot of flux
ratio or magnitude difference vs.\ binary star separation.  The data
points mark the detected companion stars, the thick line shows the
average and the thin line the worst sensitivity for undetected
companions.  The dashed vertical line at $0.13''$ shows the
diffraction limit for a $3.5\rm\,m$ telescope at K.  This is the limit
for unambiguous identification of binary stars.  The dashed horizontal
line shows the completeness limit in flux ratio for the whole survey.
The two observations of RXJ~1039.5-7538 are represented by two data
points connected by a line.}
\label{MaxBrgtAllFig}
\end{figure*}

If the object appears unresolved, we compute the maximum brightness
ratio of a companion that could be hidden in the noise of the data.
The principle is to determine how far the data deviate from the
nominal result for a point source (modulus${}=1$, phase${}=0$).  We
then compute the maximum brightness ratio of a companion that would
be compatible with this amount of deviation.  This is repeated for
different position angles, and the maximum is used as upper limit for
the brightness ratio of an undetected companion. See
\citet{LeinertHenry97} for a more detailed description of this
procedure.

If the object is a binary or triple, we compute a multidimensional
least-squares fit using the {\tt amoeba} algorithm \citep{Press94}.
Our program tries to minimize the difference between modulus and phase
computed from a model binary or triple and the observational data by
varying the separation, position angle, and brightness ratio of the
model.  This is necessary because the reconstructed images are a complex
function of the 2-dimensional separation vector and flux ratio that
cannot be solved to compute the binary parameters directly from the
data.  Fits to different subsets of the data yield an estimate for the
standard deviation of the binary parameters.  We then subtract the
contribution of the companion(s) from the images and apply the
procedure described in the previous paragraph to find limits for the
brightness of an undetected companion.

In order to find binaries that are separated by more than $3''$, we
obtained additional infrared images with the ESO/\allowbreak MPIA
$2.2\rm\,m$ telescope on La Silla in February and March 1996 using the
IRAC2b camera.  For data reduction, we used the Daophot package within
IRAF\footnote{IRAF is distributed by the National Optical Astronomy
Observatories, which are operated by the Association of Universities
for Research in Astronomy, Inc., under cooperative agreement with the
National Science Foundation.}.


\section{Results}

\subsection{Uncorrected Data}

All the binary and multiple stars we find in our sample are listed in
Table~\ref{BinTab}, while Table~\ref{SingTab}
gives limits for the brightness of undetected companions.
Figure~\ref{MaxBrgtAllFig} shows these results as a plot of flux ratio
and magnitude difference vs.\ binary separation.  In total, among the
81 systems of the sample we find 18 binaries and one triple star.

The modulus of the complex visibility of a binary is a cosine-shaped
function.  If the separation of the binary is equal to the diffraction
limit ($0.13''$ for a $3.5\rm\,m$ telescope at K), exactly one period
of the modulus of the visibility fits within the radius where the
optical transfer function of the telescope is not zero.  Under good
circumstances, it is possible to discover binaries with even smaller
separations, down to about half the diffraction limit
(Table~\ref{BinTab} shows that we actually do find some, an example is
presented in Figure~\ref{ExmplFig}).  However, in these cases we
can detect only the first minimum, but not the second maximum of the
modulus of the visibility. Therefore, we cannot distinguish with
certainty a close binary star from an elongated structure.  Even more
important, we cannot be sure that we find {\em all\/} companions at
separations less than the diffraction limit.  For these reasons, we
limit ourselves to companions in the separation range between $0.13''$
and $6''$.  The upper limit was chosen so that contamination with
background stars has little effect (see section \ref{bgsect} for a
detailed discussion of this problem).

We found four binaries with separations smaller than $0.13''$.  These
binaries are treated as unresolved stars in our survey.  This leaves
14 binaries and one triple, i.\,e.\ 16 companion stars, including one
binary in the $\eta$ Cha cluster.  For ease of comparison we mention
that these uncorrected data correspond to a fractional multiplicity
(number of multiples divided by total number of systems) of $0.19\pm
0.05$ or to a number of $0.20\pm 0.05$ companions per primary.  These
numbers change to $0.18\pm 0.05$ and $0.19\pm 0.05$ if we exclude the
four stars in the $\eta$ Chamaeleontis cluster.

\newpage
\subsection{Completeness}
\label{complsect}
Figure~\ref{MaxBrgtAllFig} shows not only the stars where we find
companions, but also the sensitivity of our survey, i.\,e.\ the
maximum brightness ratio of a possible undetected companion as a
function of the separation.  This sensitivity depends on factors like
the atmospheric conditions at the time of the observations and the
brightness of the target star.  Fig.~\ref{MaxBrgtAllFig} and
Table~\ref{SingTab} show that in most cases we are able to detect
companions with magnitude differences to the primary up to (at least)
$2.5^{\rm mag}$.  This corresponds to a brightness ratio of 1:10.

Our observations of four stars are not sufficiently sensitive to
exclude faint companions at small separations.  However, based on the
number of companions actually found, we expect about 0.3 additional
companions above a brightness ratio of 1:10 at separations ${}>
0.13''$.  Therefore, we are confident we have found all companions
with separations between $0.13''$ and $6''$ and with a magnitude
difference of less than $2.5^{\rm mag}$.

\begin{deluxetable}{rl@{}lcr *{2}{r@{${}\pm $}l} l@{${}\pm $}l}
\tablecaption{Binary and triple stars.%
\label{BinTab}}
%
%
\tablehead{%
\,No.\hfill & Designation && Evol. &
	\multicolumn{1}{c}{Date of}&
	\omit\hfil Separation\span\omit\hfil &
	\omit\hfil Position\span\omit\hfil&
	\omit\hfil Brightness\span\omit\hfil \\
	&		&& Status\tablenotemark{a} &
	\multicolumn{1}{c}{Observation} &
	\omit\hfil [$"$]\span\omit\hfil &
	\omit\hfil Angle [$^\circ$]\span\omit\hfil&
	\omit\hfil Ratio at K\span\omit\hfil }
\startdata
1 & RXJ 0837.0-7856\tablenotemark{b} &  & PMS & 29.\,Mar.\,96~ & 0.135 & 0.003 & 15.9 & 1.8 & 0.805 & 0.024 \\
11 & RXJ 0915.5-7609 &  & PMS & 30.\,Mar.\,96~ & 0.111 & 0.007 & 292.5 & 4.3 & 0.255 & 0.02 \\
13 & RXJ 0919.4-7738N &  & ? & 30.\,Mar.\,96~ & 0.109 & 0.003 & 173.9 & 1.2 & 0.658 & 0.021 \\
16 & RXJ 0933.2-7433 &  & ZAMS & 31.\,Mar.\,96~ & 0.222 & 0.003 & 231.4 & 0.4 & 0.743 & 0.034 \\
17 & RXJ 0935.0-7804 &  & PMS & 31.\,Mar.\,96~ & 0.360 & 0.003 & 353.9 & 0.2 & 0.901 & 0.022 \\
22 & RXJ 0952.7-7933 &  & ? & 31.\,Mar.\,96~ & 0.267 & 0.003 & 314.6 & 0.5 & 0.259 & 0.018 \\
26 & RXJ 1009.6-8105 &  & ? & 28.\,Feb.\,96~ & 5.413 & 0.017 & 136.9 & 0.3 & 0.039 & 0.001 \\
27 & RXJ 1014.2-7636 &  & ? & 1.\,Apr.\,96~ & 0.091 & 0.007 & 259.6 & 6.5 & 0.593 & 0.062 \\
31/32 & RXJ 1039.5-7538N+S &  & ? & 26.\,Feb.\,96~ & 4.962 & 0.022 & 347.4 & 0.2 & 0.287 & 0.003 \\
 &  &  &  & 30.\,Mar.\,96~ & 5.002 & 0.003 & 347.5 & 0.1 & 0.264 & 0.004 \\
36 & RXJ 1108.8-7519b &  & PMS & 30.\,Mar.\,96~ & 0.150 & 0.003 & 7.8 & 1.7 & 0.716 & 0.019 \\
43 & RXJ 1125.8-8456 &  & ? & 26.\,Feb.\,96~ & 3.545 & 0.022 & 208.0 & 0.2 & 0.140 & 0.003 \\
44 & RXJ 1129.2-7546 &  & PMS & 29.\,Mar.\,96~ & 0.534 & 0.003 & 28.7 & 0.2 & 0.189 & 0.008 \\
48 & RXJ 1150.9-7411 &  & ? & 1.\,Apr.\,96~ & 0.875 & 0.003 & 106.0 & 0.1 & 0.257 & 0.012 \\
50 & RXJ 1158.5-7754a &  & PMS & 29.\,Mar.\,96~ & 0.073 & 0.006 & 160.8 & 2.8 & 0.506 & 0.089 \\
57 & RXJ 1203.7-8129 &  & ? & 31.\,Mar.\,96~ & 0.396 & 0.003 & 238.3 & 0.1 & 0.775 & 0.008 \\
60 & RXJ 1207.9-7555 &  & ? & 1.\,Apr.\,96~ & 0.588 & 0.003 & 10.2 & 0.2 & 0.924 & 0.02 \\
65 & RXJ 1220.4-7407 &  & PMS & 1.\,Apr.\,96~ & 0.296 & 0.003 & 348.4 & 1.2 & 0.258 & 0.021 \\
73 & RXJ 1243.1-7458 & A-B & PMS & 30.\,Mar.\,96~ & 0.280 & 0.004 & 84.7 & 3.6 & 0.386 & 0.025 \\
 &  & AB-C &  & 30.\,Mar.\,96~ & 2.513 & 0.003 & 258.7 & 0.1 & 0.244 & 0.007 \\
75 & RXJ 1301.0-7654 &  & PMS & 30.\,Mar.\,96~ & 1.444 & 0.005 & 3.2 & 0.2 & 0.155 & 0.01 \\
\enddata
\tablenotetext{a}{\citet{Covino97}}
\tablenotetext{b}{member of the $\eta$ Chamaeleontis Cluster \citep{Mamajek99}}
\end{deluxetable}
\begin{deluxetable}{rlcr cccc}
\tablecaption{Limits for the brightness of undetected companions.%
\label{SingTab}}
%
%
\tablehead{%
No. & Designation & Evol.		  & \colhead{Date of} &
	\multicolumn{2}{c}{Maximum Flux Ratio} &
	\multicolumn{2}{c}{Minimal $\Delta m_{\rm K}$ [mag]} \\
    &		& Status\tablenotemark{a} & \colhead{Observation} &
	~at $0.13''$ & at $0.5''$	& ~at $0.13''$ & at $0.5''$}
\startdata
1 & \objectname{RXJ 0837.0-7856}\tablenotemark{b} & PMS & 29.\,Mar.\,96~ & \hbox to2em{0.04\hss} & \hbox to2em{0.01\hss} & \hbox to2em{3.49\hss} & \hbox to2em{5.00\hss}\\
2 & \objectname{RXJ 0842.4-8345} & ? & 30.\,Mar.\,96~ & \hbox to2em{0.05\hss} & \hbox to2em{0.02\hss} & \hbox to2em{3.25\hss} & \hbox to2em{4.25\hss}\\
3 & \objectname{RXJ 0842.9-7904}\tablenotemark{b} & PMS & 29.\,Mar.\,96~ & \hbox to2em{0.05\hss} & \hbox to2em{0.01\hss} & \hbox to2em{3.25\hss} & \hbox to2em{5.00\hss}\\
4 & \objectname{RXJ 0844.5-7846}\tablenotemark{b} & PMS & 29.\,Mar.\,96~ & \hbox to2em{0.05\hss} & \hbox to2em{0.02\hss} & \hbox to2em{3.25\hss} & \hbox to2em{4.25\hss}\\
5 & \objectname{RXJ 0848.0-7854}\tablenotemark{b} & PMS & 29.\,Mar.\,96~ & \hbox to2em{0.05\hss} & \hbox to2em{0.02\hss} & \hbox to2em{3.25\hss} & \hbox to2em{4.25\hss}\\
6 & \objectname{RXJ 0849.2-7735} & ? & 29.\,Mar.\,96~ & \hbox to2em{0.04\hss} & \hbox to2em{0.01\hss} & \hbox to2em{3.49\hss} & \hbox to2em{5.00\hss}\\
7 & \objectname{RXJ 0850.1-7554} & PMS & 29.\,Mar.\,96~ & \hbox to2em{0.06\hss} & \hbox to2em{0.01\hss} & \hbox to2em{3.05\hss} & \hbox to2em{5.00\hss}\\
8 & \objectname{RXJ 0853.1-8244} & ? & 30.\,Mar.\,96~ & \hbox to2em{0.05\hss} & \hbox to2em{0.02\hss} & \hbox to2em{3.25\hss} & \hbox to2em{4.25\hss}\\
9 & \objectname{RXJ 0901.0-7715} & ? & 30.\,Mar.\,96~ & \hbox to2em{0.2\hss} & \hbox to2em{0.07\hss} & \hbox to2em{1.75\hss} & \hbox to2em{2.89\hss}\\
10 & \objectname{RXJ 0902.9-7759} & PMS & 30.\,Mar.\,96~ & \hbox to2em{0.05\hss} & \hbox to2em{0.02\hss} & \hbox to2em{3.25\hss} & \hbox to2em{4.25\hss}\\
11 & \objectname{RXJ 0915.5-7609} & PMS & 30.\,Mar.\,96~ & \hbox to2em{0.05\hss} & \hbox to2em{0.01\hss} & \hbox to2em{3.25\hss} & \hbox to2em{5.00\hss}\\
12 & \objectname{RXJ 0917.2-7744} & ? & 30.\,Mar.\,96~ & \hbox to2em{0.07\hss} & \hbox to2em{0.01\hss} & \hbox to2em{2.89\hss} & \hbox to2em{5.00\hss}\\
13 & \objectname{RXJ 0919.4-7738N} & ? & 30.\,Mar.\,96~ & \hbox to2em{0.04\hss} & \hbox to2em{0.01\hss} & \hbox to2em{3.49\hss} & \hbox to2em{5.00\hss}\\
14 & \objectname{RXJ 0919.4-7738S} & ? & 30.\,Mar.\,96~ & \hbox to2em{0.17\hss} & \hbox to2em{0.07\hss} & \hbox to2em{1.92\hss} & \hbox to2em{2.89\hss}\\
15 & \objectname{RXJ 0928.5-7815} & ? & 30.\,Mar.\,96~ & \hbox to2em{0.08\hss} & \hbox to2em{0.02\hss} & \hbox to2em{2.74\hss} & \hbox to2em{4.25\hss}\\
16 & \objectname{RXJ 0933.2-7433} & ZAMS & 31.\,Mar.\,96~ & \hbox to2em{0.06\hss} & \hbox to2em{0.01\hss} & \hbox to2em{3.05\hss} & \hbox to2em{5.00\hss}\\
17 & \objectname{RXJ 0935.0-7804} & PMS & 31.\,Mar.\,96~ & \hbox to2em{0.04\hss} & \hbox to2em{0.01\hss} & \hbox to2em{3.49\hss} & \hbox to2em{5.00\hss}\\
18 & \objectname{RXJ 0936.3-7820} & ? & 31.\,Mar.\,96~ & \hbox to2em{0.06\hss} & \hbox to2em{0.02\hss} & \hbox to2em{3.05\hss} & \hbox to2em{4.25\hss}\\
19 & \objectname{RXJ 0942.7-7726} & PMS & 31.\,Mar.\,96~ & \hbox to2em{0.03\hss} & \hbox to2em{0.01\hss} & \hbox to2em{3.81\hss} & \hbox to2em{5.00\hss}\\
20 & \objectname{RXJ 0946.9-8011} & ? & 31.\,Mar.\,96~ & \hbox to2em{0.05\hss} & \hbox to2em{0.03\hss} & \hbox to2em{3.25\hss} & \hbox to2em{3.81\hss}\\
21 & \objectname{RXJ 0951.9-7901} & PMS & 31.\,Mar.\,96~ & \hbox to2em{0.04\hss} & \hbox to2em{0.02\hss} & \hbox to2em{3.49\hss} & \hbox to2em{4.25\hss}\\
22 & \objectname{RXJ 0952.7-7933} & ? & 31.\,Mar.\,96~ & \hbox to2em{0.05\hss} & \hbox to2em{0.01\hss} & \hbox to2em{3.25\hss} & \hbox to2em{5.00\hss}\\
23 & \objectname{RXJ 1001.1-7913} & PMS & 31.\,Mar.\,96~ & \hbox to2em{0.05\hss} & \hbox to2em{0.01\hss} & \hbox to2em{3.25\hss} & \hbox to2em{5.00\hss}\\
24 & \objectname{RXJ 1005.3-7749} & PMS & 1.\,Apr.\,96~ & \hbox to2em{0.06\hss} & \hbox to2em{0.01\hss} & \hbox to2em{3.05\hss} & \hbox to2em{5.00\hss}\\
25 & \objectname{RXJ 1007.7-8504} & ? & 1.\,Apr.\,96~ & \hbox to2em{0.05\hss} & \hbox to2em{0.01\hss} & \hbox to2em{3.25\hss} & \hbox to2em{5.00\hss}\\
26 & \objectname{RXJ 1009.6-8105} & ? & 1.\,Apr.\,96~ & \hbox to2em{0.06\hss} & \hbox to2em{0.02\hss} & \hbox to2em{3.05\hss} & \hbox to2em{4.25\hss}\\
27 & \objectname{RXJ 1014.2-7636} & ? & 1.\,Apr.\,96~ & \hbox to2em{0.05\hss} & \hbox to2em{0.01\hss} & \hbox to2em{3.25\hss} & \hbox to2em{5.00\hss}\\
28 & \objectname{RXJ 1014.4-8138} & ? & 1.\,Apr.\,96~ & \hbox to2em{0.04\hss} & \hbox to2em{0.02\hss} & \hbox to2em{3.49\hss} & \hbox to2em{4.25\hss}\\
29 & \objectname{RXJ 1017.9-7431} & ZAMS & 1.\,Apr.\,96~ & \hbox to2em{0.05\hss} & \hbox to2em{0.02\hss} & \hbox to2em{3.25\hss} & \hbox to2em{4.25\hss}\\
30 & \objectname{RXJ 1035.8-7859} & ZAMS & 30.\,Mar.\,96~ & \hbox to2em{0.06\hss} & \hbox to2em{0.03\hss} & \hbox to2em{3.05\hss} & \hbox to2em{3.81\hss}\\
31/32 & \objectname{RXJ 1039.5-7538N+S} & ? & 30.\,Mar.\,96~ & \hbox to2em{0.12\hss} & \hbox to2em{0.05\hss} & \hbox to2em{2.30\hss} & \hbox to2em{3.25\hss}\\
33 & \objectname{RXJ 1044.6-7849} & ZAMS & 30.\,Mar.\,96~ & \hbox to2em{0.10\hss} & \hbox to2em{0.02\hss} & \hbox to2em{2.50\hss} & \hbox to2em{4.25\hss}\\
34 & \objectname{RXJ 1048.9-7655} & ? & 30.\,Mar.\,96~ & \hbox to2em{0.06\hss} & \hbox to2em{0.03\hss} & \hbox to2em{3.05\hss} & \hbox to2em{3.81\hss}\\
35 & \objectname{RXJ 1108.8-7519a} & PMS & 30.\,Mar.\,96~ & \hbox to2em{0.13\hss} & \hbox to2em{0.04\hss} & \hbox to2em{2.22\hss} & \hbox to2em{3.49\hss}\\
36 & \objectname{RXJ 1108.8-7519b} & PMS & 30.\,Mar.\,96~ & \hbox to2em{0.05\hss} & \hbox to2em{0.01\hss} & \hbox to2em{3.25\hss} & \hbox to2em{5.00\hss}\\
37 & \objectname{RXJ 1109.4-7627} & PMS & 30.\,Mar.\,96~ & \hbox to2em{0.09\hss} & \hbox to2em{0.02\hss} & \hbox to2em{2.61\hss} & \hbox to2em{4.25\hss}\\
38 & \objectname{RXJ 1111.7-7620} & PMS & 30.\,Mar.\,96~ & \hbox to2em{0.06\hss} & \hbox to2em{0.03\hss} & \hbox to2em{3.05\hss} & \hbox to2em{3.81\hss}\\
39 & \objectname{RXJ 1112.7-7637} & PMS & 29.\,Mar.\,96~ & \hbox to2em{0.07\hss} & \hbox to2em{0.05\hss} & \hbox to2em{2.89\hss} & \hbox to2em{3.25\hss}\\
40 & \objectname{RXJ 1117.0-8028} & PMS & 30.\,Mar.\,96~ & \hbox to2em{0.05\hss} & \hbox to2em{0.03\hss} & \hbox to2em{3.25\hss} & \hbox to2em{3.81\hss}\\
41 & \objectname{RXJ 1120.3-7828} & ? & 29.\,Mar.\,96~ & \hbox to2em{0.04\hss} & \hbox to2em{0.02\hss} & \hbox to2em{3.49\hss} & \hbox to2em{4.25\hss}\\
42 & \objectname{RXJ 1123.2-7924} & PMS & 30.\,Mar.\,96~ & \hbox to2em{0.07\hss} & \hbox to2em{0.05\hss} & \hbox to2em{2.89\hss} & \hbox to2em{3.25\hss}\\
43 & \objectname{RXJ 1125.8-8456} & ? & 30.\,Mar.\,96~ & \hbox to2em{0.05\hss} & \hbox to2em{0.02\hss} & \hbox to2em{3.25\hss} & \hbox to2em{4.25\hss}\\
44 & \objectname{RXJ 1129.2-7546} & PMS & 29.\,Mar.\,96~ & \hbox to2em{0.05\hss} & \hbox to2em{0.01\hss} & \hbox to2em{3.25\hss} & \hbox to2em{5.00\hss}\\
45 & \objectname{RXJ 1140.3-8321} & ? & 30.\,Mar.\,96~ & \hbox to2em{0.03\hss} & \hbox to2em{0.03\hss} & \hbox to2em{3.81\hss} & \hbox to2em{3.81\hss}\\
46 & \objectname{RXJ 1149.8-7850} & PMS & 1.\,Apr.\,96~ & \hbox to2em{0.04\hss} & \hbox to2em{0.03\hss} & \hbox to2em{3.49\hss} & \hbox to2em{3.81\hss}\\
47 & \objectname{RXJ 1150.4-7704} & PMS & 29.\,Mar.\,96~ & \hbox to2em{0.10\hss} & \hbox to2em{0.02\hss} & \hbox to2em{2.50\hss} & \hbox to2em{4.25\hss}\\
48 & \objectname{RXJ 1150.9-7411} & ? & 1.\,Apr.\,96~ & \hbox to2em{0.08\hss} & \hbox to2em{0.01\hss} & \hbox to2em{2.74\hss} & \hbox to2em{5.00\hss}\\
49 & \objectname{RXJ 1157.2-7921} & PMS & 29.\,Mar.\,96~ & \hbox to2em{0.10\hss} & \hbox to2em{0.03\hss} & \hbox to2em{2.50\hss} & \hbox to2em{3.81\hss}\\
50 & \objectname{RXJ 1158.5-7754a} & PMS & 29.\,Mar.\,96~ & \hbox to2em{0.03\hss} & \hbox to2em{0.01\hss} & \hbox to2em{3.81\hss} & \hbox to2em{5.00\hss}\\
51 & \objectname{RXJ 1158.5-7754b} & PMS & 29.\,Mar.\,96~ & \hbox to2em{0.05\hss} & \hbox to2em{0.03\hss} & \hbox to2em{3.25\hss} & \hbox to2em{3.81\hss}\\
52 & \objectname{RXJ 1158.5-7913} & PMS & 29.\,Mar.\,96~ & \hbox to2em{0.06\hss} & \hbox to2em{0.01\hss} & \hbox to2em{3.05\hss} & \hbox to2em{5.00\hss}\\
53 & \objectname{RXJ 1159.7-7601} & PMS & 31.\,Mar.\,96~ & \hbox to2em{0.02\hss} & \hbox to2em{0.01\hss} & \hbox to2em{4.25\hss} & \hbox to2em{5.00\hss}\\
54 & \objectname{RXJ 1201.7-7859} & PMS & 31.\,Mar.\,96~ & \hbox to2em{0.06\hss} & \hbox to2em{0.01\hss} & \hbox to2em{3.05\hss} & \hbox to2em{5.00\hss}\\
55 & \objectname{RXJ 1202.1-7853} & PMS & 31.\,Mar.\,96~ & \hbox to2em{0.04\hss} & \hbox to2em{0.01\hss} & \hbox to2em{3.49\hss} & \hbox to2em{5.00\hss}\\
56 & \objectname{RXJ 1202.8-7718} & ? & 31.\,Mar.\,96~ & \hbox to2em{0.03\hss} & \hbox to2em{0.01\hss} & \hbox to2em{3.81\hss} & \hbox to2em{5.00\hss}\\
57 & \objectname{RXJ 1203.7-8129} & ? & 31.\,Mar.\,96~ & \hbox to2em{0.03\hss} & \hbox to2em{0.01\hss} & \hbox to2em{3.81\hss} & \hbox to2em{5.00\hss}\\
58 & \objectname{RXJ 1204.6-7731} & PMS & 31.\,Mar.\,96~ & \hbox to2em{0.04\hss} & \hbox to2em{0.01\hss} & \hbox to2em{3.49\hss} & \hbox to2em{5.00\hss}\\
59 & \objectname{RXJ 1207.7-7953} & ? & 1.\,Apr.\,96~ & \hbox to2em{0.07\hss} & \hbox to2em{0.03\hss} & \hbox to2em{2.89\hss} & \hbox to2em{3.81\hss}\\
60 & \objectname{RXJ 1207.9-7555} & ? & 1.\,Apr.\,96~ & \hbox to2em{0.04\hss} & \hbox to2em{0.01\hss} & \hbox to2em{3.49\hss} & \hbox to2em{5.00\hss}\\
61 & \objectname{RXJ 1209.8-7344} & ? & 1.\,Apr.\,96~ & \hbox to2em{0.04\hss} & \hbox to2em{0.02\hss} & \hbox to2em{3.49\hss} & \hbox to2em{4.25\hss}\\
62 & \objectname{RXJ 1216.8-7753} & PMS & 1.\,Apr.\,96~ & \hbox to2em{0.10\hss} & \hbox to2em{0.04\hss} & \hbox to2em{2.50\hss} & \hbox to2em{3.49\hss}\\
63 & \objectname{RXJ 1217.4-8035} & ? & 1.\,Apr.\,96~ & \hbox to2em{0.07\hss} & \hbox to2em{0.01\hss} & \hbox to2em{2.89\hss} & \hbox to2em{5.00\hss}\\
64 & \objectname{RXJ 1219.7-7403} & PMS & 1.\,Apr.\,96~ & \hbox to2em{0.04\hss} & \hbox to2em{0.02\hss} & \hbox to2em{3.49\hss} & \hbox to2em{4.25\hss}\\
65 & \objectname{RXJ 1220.4-7407} & PMS & 1.\,Apr.\,96~ & \hbox to2em{0.05\hss} & \hbox to2em{0.01\hss} & \hbox to2em{3.25\hss} & \hbox to2em{5.00\hss}\\
66 & \objectname{RXJ 1220.6-7539} & ? & 1.\,Apr.\,96~ & \hbox to2em{0.03\hss} & \hbox to2em{0.01\hss} & \hbox to2em{3.81\hss} & \hbox to2em{5.00\hss}\\
67 & \objectname{RXJ 1223.5-7740} & ? & 1.\,Apr.\,96~ & \hbox to2em{0.04\hss} & \hbox to2em{0.02\hss} & \hbox to2em{3.49\hss} & \hbox to2em{4.25\hss}\\
68 & \objectname{RXJ 1224.8-7503} & ? & 1.\,Apr.\,96~ & \hbox to2em{0.04\hss} & \hbox to2em{0.02\hss} & \hbox to2em{3.49\hss} & \hbox to2em{4.25\hss}\\
69 & \objectname{RXJ 1225.3-7857} & ? & 1.\,Apr.\,96~ & \hbox to2em{0.08\hss} & \hbox to2em{0.01\hss} & \hbox to2em{2.74\hss} & \hbox to2em{5.00\hss}\\
70 & \objectname{RXJ 1231.9-7848} & ? & 1.\,Apr.\,96~ & \hbox to2em{0.06\hss} & \hbox to2em{0.03\hss} & \hbox to2em{3.05\hss} & \hbox to2em{3.81\hss}\\
71 & \objectname{RXJ 1233.5-7523} & ? & 30.\,Mar.\,96~ & \hbox to2em{0.04\hss} & \hbox to2em{0.02\hss} & \hbox to2em{3.49\hss} & \hbox to2em{4.25\hss}\\
72 & \objectname{RXJ 1239.4-7502} & PMS & 29.\,Mar.\,96~ & \hbox to2em{0.05\hss} & \hbox to2em{0.01\hss} & \hbox to2em{3.25\hss} & \hbox to2em{5.00\hss}\\
73 & \objectname{RXJ 1243.1-7458} & PMS & 30.\,Mar.\,96~ & \hbox to2em{0.1\hss} & \hbox to2em{0.02\hss} & \hbox to2em{2.50\hss} & \hbox to2em{4.25\hss}\\
74 & \objectname{RXJ 1243.6-7834} & ? & 30.\,Mar.\,96~ & \hbox to2em{0.10\hss} & \hbox to2em{0.03\hss} & \hbox to2em{2.50\hss} & \hbox to2em{3.81\hss}\\
75 & \objectname{RXJ 1301.0-7654} & PMS & 30.\,Mar.\,96~ & \hbox to2em{0.09\hss} & \hbox to2em{0.02\hss} & \hbox to2em{2.61\hss} & \hbox to2em{4.25\hss}\\
76 & \objectname{RXJ 1303.3-7706} & ? & 30.\,Mar.\,96~ & \hbox to2em{0.08\hss} & \hbox to2em{0.04\hss} & \hbox to2em{2.74\hss} & \hbox to2em{3.49\hss}\\
77 & \objectname{RXJ 1307.3-7602} & ? & 30.\,Mar.\,96~ & \hbox to2em{0.05\hss} & \hbox to2em{0.04\hss} & \hbox to2em{3.25\hss} & \hbox to2em{3.49\hss}\\
78 & \objectname{RXJ 1320.0-7406} & ? & 31.\,Mar.\,96~ & \hbox to2em{0.11\hss} & \hbox to2em{0.05\hss} & \hbox to2em{2.40\hss} & \hbox to2em{3.25\hss}\\
79 & \objectname{RXJ 1325.7-7955} & ? & 31.\,Mar.\,96~ & \hbox to2em{0.06\hss} & \hbox to2em{0.01\hss} & \hbox to2em{3.05\hss} & \hbox to2em{5.00\hss}\\
80 & \objectname{RXJ 1346.4-7409} & ? & 31.\,Mar.\,96~ & \hbox to2em{0.12\hss} & \hbox to2em{0.04\hss} & \hbox to2em{2.30\hss} & \hbox to2em{3.49\hss}\\
81 & \objectname{RXJ 1349.2-7549} & ? & 1.\,Apr.\,96~ & \hbox to2em{0.03\hss} & \hbox to2em{0.01\hss} & \hbox to2em{3.81\hss} & \hbox to2em{5.00\hss}\\
82 & \objectname{RXJ 1415.0-7822} & PMS & 1.\,Apr.\,96~ & \hbox to2em{0.06\hss} & \hbox to2em{0.01\hss} & \hbox to2em{3.05\hss} & \hbox to2em{5.00\hss}\\
\enddata
\tablenotetext{a}{\citet{Covino97}}
\tablenotetext{b}{member of the $\eta$ Chamaeleontis cluster \citep{Mamajek99}}
\end{deluxetable}
\clearpage

\subsection{Confusion with Background Stars}
\label{bgsect}

Since we observed our targets only once, we have no possibility to
detect orbital motion.  Therefore, we cannot say if a given binary is
indeed a physically bound pair or only a chance alignment with a field
star.  To estimate the number of chance projections, we count the
field stars in 76 of the infrared images obtained at the
ESO/\allowbreak MPIA $2.2\rm\,m$ telescope.  We exclude a circular
area with a radius of $15''$ around the T~Tauri star in each image to
avoid counting physically bound companions.  This leaves $4400\,\rm
arcsec^2$ per field, and a total area of $93\,\rm arcmin^2$.

\begin{figure}[t]
\centerline{\psfig{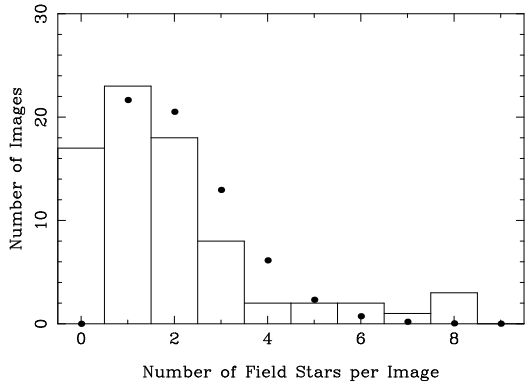}}
\caption{Distribution of the number of field stars in 76 images.
The histogram shows the number of background stars we count in our
images; the dots denote a Poisson distribution with the same average
star density}
\label{BgDistFig}
\end{figure}

The distribution of the number of field stars we obtain in this way is
shown in Fig.~\ref{BgDistFig}.  Overplotted is a Poisson distribution
with a mean of 1.89, the average number of stars found per field.  The
Poisson distribution gives a reasonable fit to the data, indicating
that the number of field stars can be described by Poisson statistics.
The average number of stars per field yields a field star density of
$(4.3\pm 0.5)\cdot 10^{-4}$ stars per $\rm arcsec^2$.  So the
probability to find a field star with a projected distance of less
than $6''$ to one T~Tauri star is
$$
	(4.3\pm 0.5)\cdot 10^{-4}\cdot \pi\cdot 6^2 = (4.9\pm 0.6)\%.
$$
The expected number of chance-projected field stars in our sample of
77 targets is
$$
	(4.9\pm 0.6)\%\cdot 77 = 3.7\pm 0.4\,.
$$

The number of physically bound companions is therefore
$15-3.7\approx11$ companion stars.  To obtain the number of
bound binaries and triples, we must take into account that the number
of ``false'' binaries depends on the number of ``true'' single stars,
and the number of ``false'' triples on the number of ``true''
binaries.  The result of this trivial, but somewhat tedious
calculation is $10.3\pm 0.4$ bound binaries and $0.5\pm 0.6$ bound
triples.  This means we cannot tell if the triple system
\objectname{RXJ~1243.1-7458} consists of three bound TTS or a binary
TTS and a field star. However, the probability for a chance projection
is proportional to the separation.  Since the pair AB-C of the triple
is one of the four widest pairs in our sample, component C is probably
a field star.

These corrected numbers correspond to a fractional multiplicity of
$(10.3+0.5)/77 = 0.140\pm 0.043$ and to $11.3/77 = 0.147\pm 0.051$
companions per primary.  The stars in the $\eta$ Chamaeleontis cluster
have been excluded to obtain this result.

\subsection{Bias Induced through X-Ray Selection}
\label{Xbiasect}

\citet{Brandner96} pointed out that the flux limit of X-ray selected
samples induces a detection bias in favour of binary and multiple
systems.  There is a small number of systems with a combined X-ray
flux above the detection threshold, although the fluxes from
individual components are below the cut-off.  These systems cause an
overestimate of the multiplicity.

\begin{figure}[t]
\centerline{\psfig{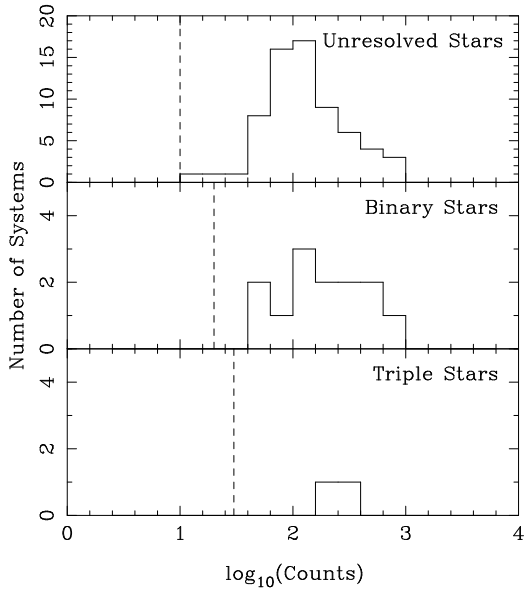}}
\caption{Distribution of the number of X-ray counts of unresolved,
binary, and triple systems.  The vertical lines mark the limits chosen
by us to obtain an unbiased sample:  $N_{\rm limit} = 10\,$counts for
unresolved stars, 20 counts for binaries, and 30 counts for triple
stars}
\label{XCntFig}
\end{figure}

To check if our results are affected by this bias, we look at the
X-ray counts that led to the detection of the stars \citep{Alcala95}
in order to search for binaries that would have been detected only
because of the bias.  These systems would have a number of counts
below twice the detection threshold, since otherwise at least one
component would be bright enough to be detected without the additional
X-ray flux from its companion.  This is a very conservative limit
since not all binaries consist of stars with equal X-ray fluxes.

The distributions of X-ray counts of unresolved, binary and triple
stars are plotted in Fig.~\ref{XCntFig}.  The lowest number of counts
measured is 12.5, while the usually adopted value for the detection
limit of the RASS is 8.  In Fig.~\ref{XCntFig}, we marked 10 counts in
the panel for unresolved stars, 20 counts in the binary panel, and 30
counts in the panel for the triple.  The figure shows that none of our
multiple systems falls even close to the detection limit.  Only two
unresolved stars could be affected if they would be binaries with
companions undetectable by our survey.  However, since the binary
fraction of our sample is below 20\,\%, we have no reason to assume
that this is the case.

We conclude that we do not have to apply any corrections for a bias
induced by the X-ray selection of our sample.

\section{Discussion}

\subsection{The Distribution of Separations of Main-Sequence Stars}
\label{MainSeqDist}

In order to compare the young binaries to main-sequence stars, we have
to face the problem that we observed the stars only once, so we know
separation and position angle at one date.
(\objectname{RXJ~1039.5-7538} was observed twice, but the timespan of
one month is not long enough to detect orbital motion.)  \citet{D+M91}
give the period distribution of main-sequence binaries, which we
cannot derive from our data.  In preceding papers
\citep[e.g.][]{Leinert93,Koehler98,Koehler2000}, we used statistical
arguments to convert the separation distribution of binaries to a
period distribution.  This relies on assumptions on the distributions
of orbital elements like eccentricity and inclination.  Furthermore,
given the small number of multiple stars in our sample, it is not
clear how reliable the result of this statistical argumentation is.

In this work, we go the other way and use the well-known distribution
of main-sequence binaries to compute their distribution of projected
separations.  To do this, we simulated samples of typically 10~million
binaries with orbital elements distributed according to \citet{D+M91},
i.e.\ the periods have a log-normal distribution, the distribution of
eccentricities is $f(e)=2e$, the inclinations are distributed
isotropically, while all the other parameters follow uniform
distributions.  In most of the simulations, we kept the system mass
fixed, but we performed simulations with different system masses
between $0.5\rm\,M_\sun$ and $2\rm\,M_\sun$.  We also simulated
samples where the system mass is not fixed, but distributed uniformly
in the ranges $0.5\ldots1.5\rm\,M_\sun$, $0.8\ldots1.5\rm\,M_\sun$,
and $0.5\ldots2\rm\,M_\sun$.  The mass ratio of the binaries does not
enter the computation, so we do not have to worry about its
distribution.

\begin{figure}[t]
\centerline{\psfig{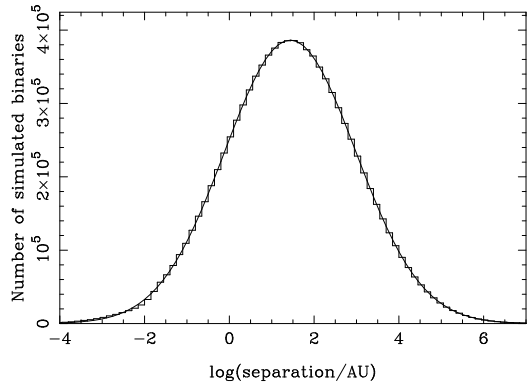}}
\caption{Simulated distribution of projected separations for a sample
of 10 million main-sequence binaries with a system mass of
$1\rm\,M_\sun$.  The histogram shows the simulated data, the line is a
log-normal distribution fitted to the histogram.}
\label{SimSepFig}
\end{figure}

Figure~\ref{SimSepFig} shows a typical result of these simulations.
The projected separations follow a log-normal distribution of the form
$$
  f(x) = C\cdot \exp(-{(x-\bar x)^2\over2\sigma^2})\,,
$$
where $x$ is the logarithm of the projected separation in AU.  To
obtain the constants $\bar x$ and $\sigma$, we fitted a Gaussian
function to the simulated data. The result for $\bar x$ depends
slightly on the system mass and varies between 1.37 and 1.51, while
all the simulations resulted in the same $\sigma$ of 1.55.  In the
following, we use $\bar x = 1.44$, the result of the simulation with
a constant system mass of $1\rm\,M_\sun$\footnote{\citet{Alcala97}
placed the ROSAT-selected stars in the Chamaeleon association in the
Hertzsprung-Russel diagram in order to determine their masses and
ages.  They found masses in the range $0.2$ -- $2.5\rm\,M_\sun$ with
about half of the stars having masses below $1\rm\,M_\sun$.  However,
they assumed a distance of 150\,pc.  If the stars are actually closer
(see Sect.~\ref{DistSect}), their real masses will be somewhat
smaller, in the range $0.15$ -- $1.7\rm\,M_\sun$}.  To get the
constant $C$, we use the number of companions given by \citet{D+M91}.
They found 101 companions in their sample of 164 main-sequence stars,
so we adjust $C$ such that the integral over $f(x)$ is 101/164.

According to the conversion from projected separation to period used
in previous papers, a separation of $10^{1.44}\rm\,AU$ corresponds to
a period of $10^{4.8}\rm\,days$, and $\sigma_{\log s} = 1.55$
translates to $\sigma_{\log P} = 2.3$.  These are the parameters of
the period distribution given by \citet{D+M91}, which demonstrates
that the old method was appropriate to compare a sample of young
binaries to main-sequence stars in the statistical sense.  However, it
gave the false impression that we are able to measure periods of
binaries based on a single observation.  Therefore, we will use the
new method in the following and compare the distributions of
separations, not periods.

\subsection{The Distance to Chamaeleon}
\label{DistSect}

The final step needed before we can compare our results to those of
\citet{D+M91} is to convert the projected separations from arcseconds
to AU.  To do so, we need the distance to the binary stars.
Hipparcos parallaxes of three stars in Cha~I give a mean distance of
$171\pm 20$\,pc \citep{Wichmann98}, in agreement with the distance of
about 140 -- 150\,pc that was obtained by various other methods (see
\citet{Schwartz91} for a review).  Hipparcos measurements of
ROSAT-discovered stars in Chamaeleon resulted in much smaller
distances between 63 and 128\,pc \citep{RNE98}.  However, four of the
seven stars observed by Hipparcos are binaries, which makes their
parallax unreliable.  The average distance of the other three stars is
65\,pc if we use the errors to weight the individual distances, and
74\,pc if we do not take the errors into account.

\citet{Frink98} studied the kinematics of T~Tauri stars in Chamaeleon,
both stars known before and stars discovered by ROSAT.  They found
that their sample can be divided into at least two subgroups, where
subgroup 1 contains mainly TTS known before ROSAT, while subgroup 2
consists mainly of ROSAT-discovered TTS.  The mean proper motion of
subgroup 2 is about twice as high as that of subgroup 1.  If both
subgroups have the same space velocity, this would imply that the mean
distance of subgroup 1 is about twice the mean distance of
subgroup~2, in agreement with the Hipparcos results.

For a given star in our sample, it is usually not possible to decide
if it belongs to the Chamaeleon association at a distance of about
140--170\,pc, or to another group in the foreground at about 70--80\,pc.
Fortunately, the difference in distance between 70 and 170\,pc
corresponds to a shift in $\log(s)$ of about 0.4, which is small
compared to the width of the separation distribution of main-sequence
stars.  In the following, we will adopt a distance of 80\,pc for the
ROSAT-selected stars, and twice as much for the stars observed by
\citet{Ghez97}.

\subsection{Comparison to Main-Sequence Stars}

At a distance of 80\,pc, the separation range 0.13 -- 6$''$
corresponds to 10.4 -- 480\,AU.  Our simulation of $1\rm\,M_\sun$
main-sequence binaries shows that $(39.6\pm 6.3)\,\%$ of the
companions are at a projected separation in this range.  We adopted an
error of $\sqrt{40}/101 = 6.3\,\%$, because 40 is 39.6\,\% of the 101
binaries found by \citet{D+M91}.  Divided by the total number of 164
systems (single stars and multiples) in the main-sequence sample, this
yields $24.1\pm 3.8$ companions per 100 systems.

In the Chamaeleon sample of 77 systems, we find $11.3\pm 3.9$
companions after correction for chance alignment with background
stars.  This corresponds to $14.7\pm 5.1$ companions per 100 systems,
which is $9.4\pm 6.3$ less than among main-sequence stars.  Thus, the
companion star frequency in our sample is reduced by a factor of
$0.61\pm 0.27$ compared to main-sequence stars.  This is surprising,
since the Chamaeleon star-forming region is a T association similar to
Taurus-Auriga, where we found a much higher binary frequency
\citep{Koehler98}.

\begin{figure}[t]
\centerline{\psfig{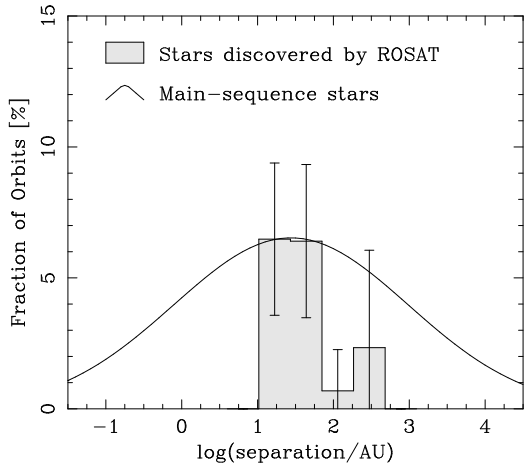}}
\caption{Companion star frequency as a function of projected
separation.  The histogram shows the result of this work, the
curve shows the distribution of separations for main-sequence
binaries.  This is the result of a simulation for system masses of
$1\rm\,M_\sun$, as described in section~\ref{MainSeqDist}.}
\label{MainSeqFig}
\end{figure}

Fig.~\ref{MainSeqFig} shows the distribution of projected separations
of the stars in our sample and main-sequence stars.  The companion
star frequencies in the range 10 -- 70\,AU agree almost perfectly, but
at larger separations, the number of companions to young stars drops
to nearly zero and is significantly lower than the number of
companions to main-sequence stars.  A Kolmogorov-Smirnov test shows
that the two distributions are indeed different: the probability that
both samples were drawn from the same distribution is only
$6\cdot 10^{-6}$.  The separation distribution of young stars in
Chamaeleon appears to be more similar to that of low-mass stars in the
Orion Trapezium Cluster: a number of binaries with small separations
that is comparable to main-sequence stars
\citep{Prosser94,Padgett97,Petr98}, and only a very small number of
binaries at larger separations \citep{Scally99}.

\begin{deluxetable}{lccccc}
\tablecaption{Companion star frequencies\label{PMSorNotTab}}
\tablehead{	& Main-sequence	& All Stars in	& PMS		  & ZAMS	& Unknown\\
	& Stars\tablenotemark{a} & Our Sample	&		  &		& Evol.~Status}
\startdata
Total		& 164		& 77		& 33		  & 4			& 40	\\
Companions\tablenotemark{b}
		&		& $11.3\pm 3.9$ & $\phn5.4\pm 2.6$& $\phn0.8\pm \phn1.0$& $\phn5.0\pm 2.6$\\
Companions per 100 systems
		& $24.1\pm 3.8$	& $14.7\pm 5.1$ & $16.3\pm 8.0$	  & $20.1\pm 25.0$	& $12.6\pm 6.6$\\
\enddata
\tablenotetext{a}{\citet{D+M91}}
\tablenotetext{b}{projected separations in the range 10.4--480\,AU,
	after subtraction of background stars}
\end{deluxetable}

\subsection{Stars of different evolutionary status}

\citet{Covino97} used high-resolution spectra to classify
ROSAT-selected stars in Chamaeleon as PMS star, ZAMS star, or star of
unknown evolutionary status.  Their object list is only a subset of
our sample; we treat stars not observed by them as stars of unknown
evolutionary status.  Table~\ref{PMSorNotTab} shows the results of our
multiplicity survey for the different classes.  Due to the large
errors caused by the small sample size, we are unable to find a
significant difference between stars of different evolutionary
status.

\begin{figure}[t]
\centerline{\psfig{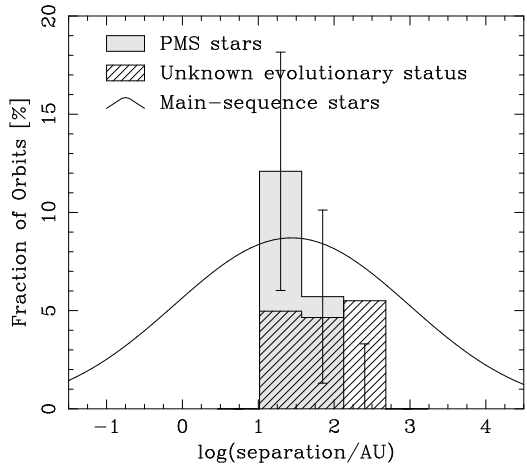}}
\caption{The distributions of projected separations of PMS stars and
stars of unknown evolutionary status.  Only error bars for PMS stars
are shown to enhance the clarity of the diagram.}
\label{PMSorNotFig}
\end{figure}

Figure~\ref{PMSorNotFig} shows the distribution of separations of PMS
stars and stars of unknown evolutionary status (it would not be useful
to plot the one ZAMS binary).  The bias towards
binaries with small separations appears to be even more pronounced
among bona-fide PMS stars, while the distribution of presumably older
stars is more or less flat.  We take this as an indication that the
unusual distribution of separations we found for the whole sample is
indeed a feature of the ROSAT-selected PMS stars in this region.

A KS test gave a probability of 38\,\% that the distributions of PMS
stars and of stars of unknown evolutionary status were drawn from the
same parent distribution.  This result does not enable us to accept or
reject the hypothesis that both distributions are identical.  The
sample of stars of unknown evolutionary status is probably a mixture
of PMS and older stars.  It shows a separation distribution that is
similar to main-sequence stars, albeit with a smaller total number of
companions.

\begin{figure}[t]
\centerline{\psfig{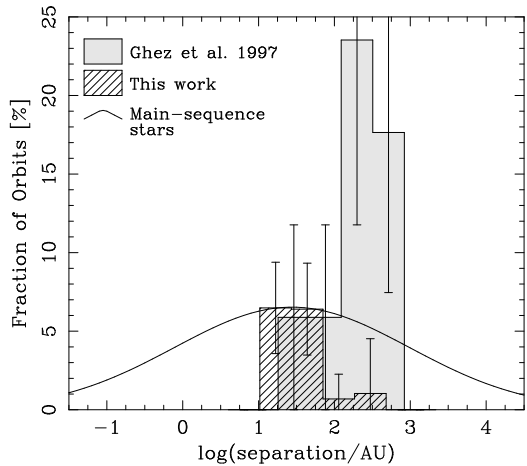}}
\caption{The distributions of projected separations of stars observed
by \citet{Ghez97} and in this work.}
\label{GhezFig}
\end{figure}

\subsection{Comparison to \citet{Ghez97}}

\citet{Ghez97} conducted a multiplicity speckle survey of TTS in
southern star-forming regions known before ROSAT, including stars in
the Chamaeleon star-forming region.  The sensitivity of their
observations for faint companions is rather non-uniform, but the
observations of 17 stars in Chamaeleon have a sensitivity comparable
to our observations.  We assumed a distance of 160\,pc for these
stars, since the majority is located in the Cha~I complex.

Figure~\ref{GhezFig} shows the distribution of projected separations
of the stars observed by \citet{Ghez97} compared to the stars observed
by us.  The distributions are clearly different.  A $\chi^2$ test
confirms this, the probability that both samples were drawn from the
same distribution is only $10^{-11}$, a KS test gave a probability of
2.4\,\%.
This difference cannot be explained by errors in the distances.
To bring the peaks of both distributions together, the distance to one
of the groups would have to be changed by a factor of 10.

This indicates that both samples represent two different populations
of stars.  This is not very surprising, since the distances of the two
groups differ by a factor of two (see section~\ref{DistSect}).
However, observations of other star-forming regions have shown that
regions with a higher stellar density contain fewer binaries
\citep{Petr98, Duchene99}.  In Chamaeleon, we find the opposite trend:
the stars observed by \citet{Ghez97} are located in the dense cores,
yet their binary frequency is lower.

There are two possible explanations for this result: either the
dependence of the binary frequency on the environmental conditions is
not a simple function of stellar density, or the ROSAT-selected stars
in Chamaeleon formed in a much denser environment.  The proper motions
of 31 of the ROSAT-selected stars are known \citep{Frink98,Frink99}.
Their proper motion vectors are not oriented away from a single point,
so they show no sign that the distribution was more concentrated in
the past.

It is more likely that the stars formed near their present location in
small ``cloudlets'' \citep{Feigelson96}.  Such small clouds have been
discovered in a ${}^{13}{\rm CO}\,(J=1$--$0)$ survey by
\citet{Mizuno98}.  They found that about 70\,\% of the TTS in the
region surveyed are located within 4\,pc of a small cloud.  However,
it is not clear why only a small number of binaries formed in the
cloudlets towards Chamaeleon, while under similar conditions in
Taurus-Auriga most stars formed in binary systems.

\section{Summary and Conclusions}

We conducted a multiplicity survey of some 80 X-ray selected young
stars.  More than 50\,\% of them have been confirmed to be
pre-main-sequence objects by measuring their Lithium abundances with
the help of high-resolution spectroscopy \citep{Covino97}.  We find a
companion star frequency that is lower by a factor of $0.61\pm 0.27$
compared to solar-type main-sequence stars \citep{D+M91}.  The
separation distributions are remarkably different: We find almost no
binaries among the young stars with separations larger than 100\,AU.

The binary frequency and separation distribution of the confirmed
pre-main-sequence stars are similar to that of the entire sample,
although with a smaller statistical significance due to the smaller
number of systems.

In contrast to our results for X-ray selected stars, \citet{Ghez97}
found a much higher binary frequency among 17 stars known before
ROSAT.  Our interpretation of this discrepancy is that the two samples
represent two different populations.  This agrees with the distance
determinations, which indicate that at least some of the
X-ray-selected stars are only half as distant as the stars known
before ROSAT.

It remains unclear how the stars formed.  According to the theory that
the multiplicity depends on the stellar density, the small number of
binaries and the shape of their separation distribution indicate that
the stars were once in a much denser environment.  However, the proper
motions show no sign of expansion from a single point.  This supports
the hypothesis that the stars formed in small cloudlets at their
present location \citep{Feigelson96,Mizuno98}, although we currently
cannot explain the bias against wide binaries.

\acknowledgements

I would like to thank Christoph Leinert for proposing this project and
for many fruitful discussions.  Andreas Eckart, Nancy Ageorges, and
Hans Zinnecker supported me during the observations with SHARP.
I modified a computer program written by Sabine Frink to carry out the
simulations described in Section~\ref{MainSeqDist}.
An anonymous referee provided a very detailed and helpful report.
This work has been supported in part by the Max-Planck-Institute for
Astronomy in Heidelberg, Germany, and by the National Science
Foundation Science and Technology Center for Adaptive Optics, managed
by the University of California at Santa Cruz under cooperative
agreement No. AST-9876783.

\end{document}